\documentclass[fleqn,usenatbib]{mnras}

\usepackage{newtxtext,newtxmath}

\usepackage[T1]{fontenc}

\DeclareRobustCommand{\VAN}[3]{#2}
\let\VANthebibliography\thebibliography
\def\thebibliography{\DeclareRobustCommand{\VAN}[3]{##3}\VANthebibliography}

\usepackage{graphicx}	% Including figure files
\usepackage{amsmath}	% Advanced maths commands
\usepackage{amssymb}	% Extra maths symbols

\title[Fe K$_{\alpha}$ line and soft X-ray lag in NGC 7456 ULX-1]{Evidence for Fe K$_{\alpha}$ line and soft X-ray lag in NGC 7456 ULX-1}

\author[S. Mondal et al.]{
Samaresh Mondal,$^{1}$\thanks{E-mail: smondal@camk.edu.pl (SM)}
Agata R{\'o}{\.z}a{\'n}ska$^{1}$,
Barbara De Marco$^{2}$
and
Alex Markowitz$^{1,3}$
\\
$^{1}$Nicolaus Copernicus Astronomical Center, Polish Academy of Sciences, ul.\ Bartycka 18, PL-00-716 Warsaw, Poland\\
$^{2}$Department de F\'{i}sica, EEBE, Universitat Polit\`{e}cnica de Catalunya, Av. Eduard Maristany 16, E-08019, Barcelona, Spain\\
$^{3}$University of California, San Diego, Center for Astrophysics and Space Sciences, MC 0424, La Jolla, CA, 92093-0424, USA
}

\date{Accepted XXX. Received YYY; in original form ZZZ}

\pubyear{2015}

\begin{document}
\label{firstpage}
\pagerange{\pageref{firstpage}--\pageref{lastpage}}
\maketitle

% Abstract of the paper
\begin{abstract}
  We report the first detection of a Fe K$_{\alpha}$ line
  and soft X-ray lag in the ultraluminous X-ray source (ULX) NGC~7456 ULX-1. 
  The \emph{XMM-Newton} spectra show the presence of the 6.4
  keV Fe line at 2.6$\sigma$ confidence and 
  an upper limit on the FWHM of
  32900 km s$^{-1}$.  Assuming that the line arises by reflection from a
  Keplerian disk, it must originate beyond $85 r_{\rm g}$ from the 
  compact object.  As a result of Fourier timing
  analysis we found that the soft X-ray photons lag behind the hard
  X-ray photons with a $\sim$1300~s delay.  The covariance spectra
  indicate that the hard spectral component is responsible for the
  correlated variability and the soft X-ray lag. This is the second
  ULX in which a Fe K$_{\alpha}$ line is found, the fifth 
  with a soft X-ray lag, and the first with both features detected. 

\end{abstract}

% Select between one and six entries from the list of approved keywords.
% Don't make up new ones.

%%%%%%%%%%%%%%%%%%%%%%%%%%%%%%%%%%%%%%%%%%%%%%%%%%

%%%%%%%%%%%%%%%%% BODY OF PAPER %%%%%%%%%%%%%%%%%%

\section{Introduction}
Ultraluminous X-ray (ULX) sources are off-nuclear point sources with
X-ray luminosity exceeding the Eddington luminosity of a 10 $\rm
M_{\odot}$ black hole (BH) ($L_{\rm X}>10^{39}\ \rm
erg\ s^{-1}$). ULXs are prime candidates in which to study
super-Eddington accretion flows, as a few of them are identified to
contain a neutron star (NS)
\citep{bachetti14,furst2016,israel2017a,israel2017b,brightman2018,carpano2018,sathyaprakash2019,rodriguez-castillo2020},
with the prospect of stellar-mass BHs existing in many others
 \citep{mondal2020b,mondal2021}. %There are various observational signatures that hint towards the presence of a radiatively-powerful thick outflow, which is a consequence of super-Eddington accretion \citep{feng2007,kajava2009,middleton2014,pinto2016,walton2016,pinto2016,kosec2018a,kosec2018b,mondal2020b}.

Most ULXs do not show strong X-ray variability,
\citep{heil2009} and those which do show it, lack short time scale ($<$ ks)
variability compared to X-ray binaries (XRBs) and active galactic
nuclei (AGN). It has been suggested that the short time scale
variability is suppressed due to the interaction of photons coming from the
inner hotter region with the outflowing material. 
X-ray variability studies made for a few ULXs, using various approaches, 
resulted in discoveries of quasi periodic oscillations
\citep{strohmayer2003}, linear rms-flux relations
\citep{heil2010,hernandez-garcia2015}, and time-lag analyses
\citep{heil2010,demarco2013}. However,
variability studies of ULXs are hampered by low count rates which
necessitate long exposure times to recover their statistical
properties.
The detection of linear rms-flux relations in NGC 5408
X-1 \citep{heil2010}, NGC 6946 X-1 \citep{hernandez-garcia2015} and
M51 ULX-7 \citep{earnshaw2016} may suggest a common origin of
X-ray variability among ULXs, XRBs, and AGNs.

\cite{walton2011} reported the presence of four ULXs in NGC 7456 based
on a 2005 \emph{XMM-Newton} observation.  Recently, \cite{pintore2020}
analyzed a deep \emph{XMM-Newton} observation from 2018 and detected an
additional ULX (ULX-5) in NGC 7456. \cite{pintore2020} performed spectral and
temporal analysis of all ULXs hosted in this galaxy and found that
ULX-1 is the brightest and most variable one. Therefore, in this paper 
we use the same \emph{XMM-Newton} observation of ULX-1, with a duration of 
$\sim$92~ks, to perform detailed studies of its X-ray variability properties. We 
report new findings obtained from spectral-timing analysis of this bright ULX.

\section{Data reduction}
We reduced the \emph{XMM-Newton} \citep{jansen2001} observation of NGC~7456
(ObsID 0824450401, 2018 May 18); its 92.4 ks duration makes it the longest 
observation for reliable spectral and timing analysis. The observation was 
processed using \emph{XMM-Newton} Science Analysis System (SASv16.0.0) following 
standard procedures. The good exposure time after screening was 82.1~ks. We only 
selected events with PATTERN$\le$4 and PATTERN$\le$12, respectively, for the EPIC-pn 
and EPIC-MOS cameras. Source spectra and lightcurves were obtained from 
a source circular region of 40$\arcsec$. Background were extracted from regions of 60$\arcsec$, 
confirmed with the \texttt{edetect\_chain} task to devoid of point sources and avoiding Cu ring on pn-CCD chip. 
The response matrices and auxiliary files were generated using the SAS tasks \texttt{rmfgen} 
and \texttt{arfgen}, respectively. 

As a first step, we used optimal spectral 
binning \citep{kaastra2016} in the aim to detect the Fe line. But it turned out that 
the continuum was binned very much up to 95 counts per bin, while the line region at 
6.4 keV had the same statistic as 20 counts per energy bin. Since we require high 
energy resolution for the continuum to constrain potential continuum curvature we use 
moderate binning. Therefore, for better constrains on the line and continuum parameters 
we use binning of 20 counts per energy bin. The lightcurves were extracted using 10~s 
time bins and background 
corrected using the SAS task \texttt{epiclccorr}. For our timing analysis we used
lightcurves from the EPIC-pn detector only as it has nearly three
times higher effective area than each MOS detector.

\section{Results}
\subsection{Time-averaged spectral analysis}
\label{sec:spec_fit}
We first perform a time-averaged spectral
fit to obtain the spectral decomposition and the unabsorbed luminosity of
the source.  \cite{pintore2020} performed detailed spectral analysis of the
same data, and found that the broad
band continuum can be fitted equally well by a number of two component
models similar to many other ULXs. In these models there is a the multi-color disk (MCD) component which peaks in
the soft X-ray band (0.3--1 keV) plus an additional component for the
hard X-ray photons, either a thermal Comptonization \citep{gladstone2009} or a hotter black body. Here
we focus on a model comprised of MCD (\texttt{diskbb})
plus thermal Comptonization   (\texttt{nthcomp}). For spectral fitting we used {\sc xspec} 
v12.11.1 \citep{arnaud1996}.

\begin{figure}
    \centering
    \includegraphics[trim={0 0 0 0},width=0.41\textwidth]{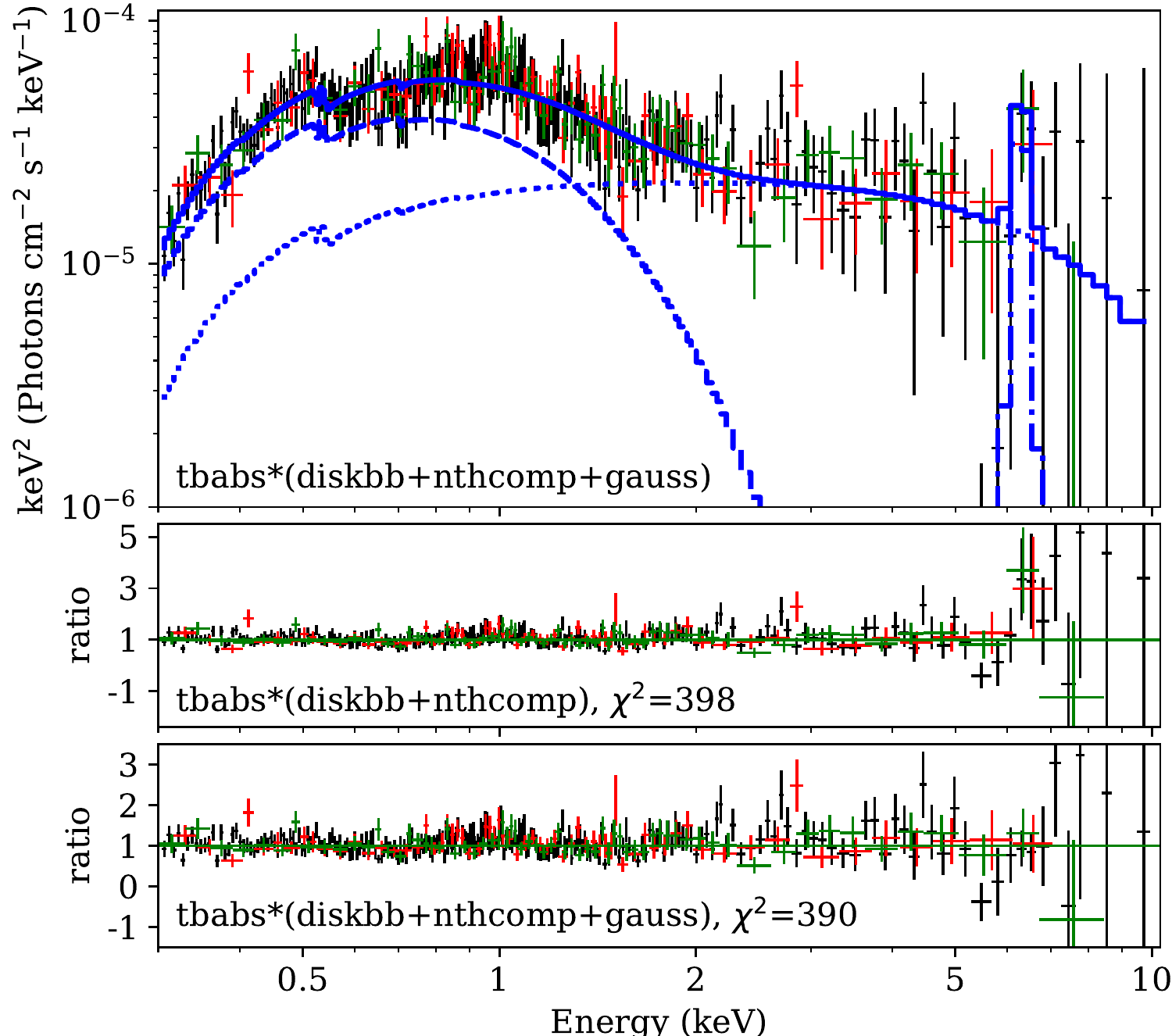}
    \caption{The top panel shows the unfolded time-averaged continuum spectrum,
      composed of a MCD (dashed curve) plus thermal Comptonization 
      (dotted curve) plus a narrow Gaussian (dotted-dash curve), while the total model is shown by the continuous line.
      Note the excess at 6.4 keV from Fe~K$_{\alpha}$ emission in the middle panel.
      The black, red and green data points are from the EPIC-pn, MOS1 and MOS2 detectors, respectively.}
    \label{fig:eeuf}
\end{figure}

Fitting with \texttt{diskbb}  plus  \texttt{nthcomp} reveals on excess near 6.4 keV 
from Fe K$_{\alpha}$ emission. This is illustrated in the data to model ratio plot
in the middle panel of Fig.~\ref{fig:eeuf}. Adding an extra Gaussian component 
to the continuum model improved the fit by |$\Delta \chi^2$|=8 for three fewer degrees of freedom. 
To estimate the statistical significance of the detection we performed Monte Carlo simulations 
using the tool \texttt{mc\_sig}\footnote{https://www.sternwarte.uni-erlangen.de/gitlab/remeis/isisscripts/-/blob/master/src/misc/simulation/mc\_sig.sl} in ISIS \citep{houck2000} 
which can take into account multiple data sets with different response files for simultaneous
analysis. The line is inconsistent with photon noise at $99.0\%$ 
confidence level (keeping all line parameters free)
or $>99.0\%$ (keeping line centroid and width $\sigma$ frozen at best-fit values). 
We also checked that the Akaike Information Criterion \cite[AIC,][]{akaike1974} and  Bayesian information criterion 
\citep[BIC,][]{schwarz1978} give lower values, after adding Gaussian component, by 
|$\Delta$AIC|=1.5 and |$\Delta$BIC|=125.

\begin{figure}
    \centering
    \includegraphics[trim={1.0cm 1.0cm 0 0},width=0.25\textwidth, angle=270]{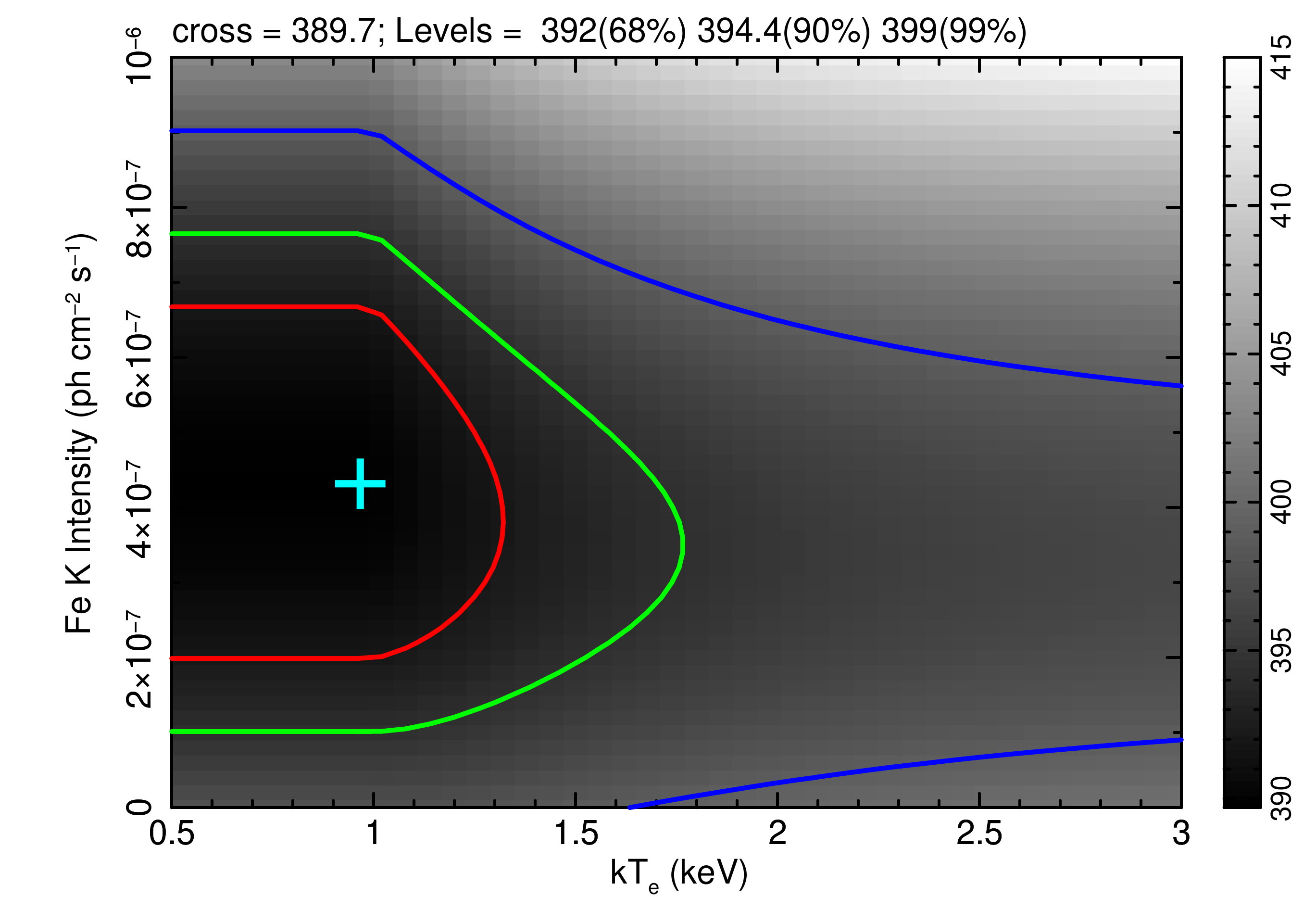}
    \caption{The $\chi^2$ contour plot of line intensity vs.\ electron temperature.
      The red, green and blue lines show the confidence levels at 68\% ($\Delta\chi^2=2.3$),
    90\% ($\Delta\chi^2=4.6$) and 99\% ($\Delta\chi^2=9.2$), respectively. The color bar shows the values of $\chi^2$.}
    \label{fig:contour}
\end{figure}

Our best-fit model for ULX-1 in NGC 7456 is
composed of an absorbed MCD plus thermal Comptonization plus a narrow
Gaussian at 6.4 keV for the Fe K$_{\alpha}$ emission:
\texttt{constant*tbabs*(diskbb+nthcomp+gauss)}, as shown in
Fig.~\ref{fig:eeuf} bottom panel.  The \texttt{constant} term is used
for cross calibration uncertainties, and we keep it free for different
detectors but fixed to unity for EPIC-pn. The fitting parameters with
90\% confidence error are: $N_{\rm H}=7.21^{+1.9}_{-1.7}\times10^{20}$ cm$^{-2}$, 
$kT_{\rm in}=0.23^{+0.03}_{-0.02}$ keV, $\Gamma=1.70^{+0.35}_{-0.24}$, 
$kT_{\rm e}=1.0\pm0.6$ keV, $E_{\rm gauss}=6.44^{+0.29}_{-0.20}$ keV, 
$\sigma=0.16\pm0.14$ keV, $N_{\rm gauss}=4.4^{+3.3}_{-2.5}\times10^{-7}$ ph~cm$^{-2}$~s$^{-1}$, and
$\chi^2/dof=389.78/373$.  The resulting equivalent width (EW) of
the line is 2000$^{+1500}_{-1100}$~eV. A contour plot of line
intensity vs.\ electron temperature of \texttt{nthcomp} model component
is shown in Fig.~\ref{fig:contour}, supporting that the line is detected at 2.6$\sigma$ significance.

The unabsorbed 0.3--10 keV flux is $2.28\times10^{-13}$ erg s$^{-1}$
cm$^{-2}$. Assuming a distance to the host galaxy of 15.7 Mpc
\citep[$z=0.00364$;][]{tully2016a}, the source unabsorbed luminosity
(0.3--10 keV) is $6.74\times10^{39}$ erg s$^{-1}$ and the line rest-frame energy 
is $(1+z)E_{\rm gauss}=6.44$ keV. The upper panel of Fig.~\ref{fig:eeuf} shows the 
spectral decomposition of the unfolded model. The soft (\texttt{diskbb})
and hard (\texttt{nthcomp}) component dominates mostly below and above
1.3 keV, respectively.

\subsection{Power spectra and fractional variability}
\label{sec:PSD}
\begin{figure}
    \centering
    \includegraphics[trim={0 0 0 0},width=0.41\textwidth]{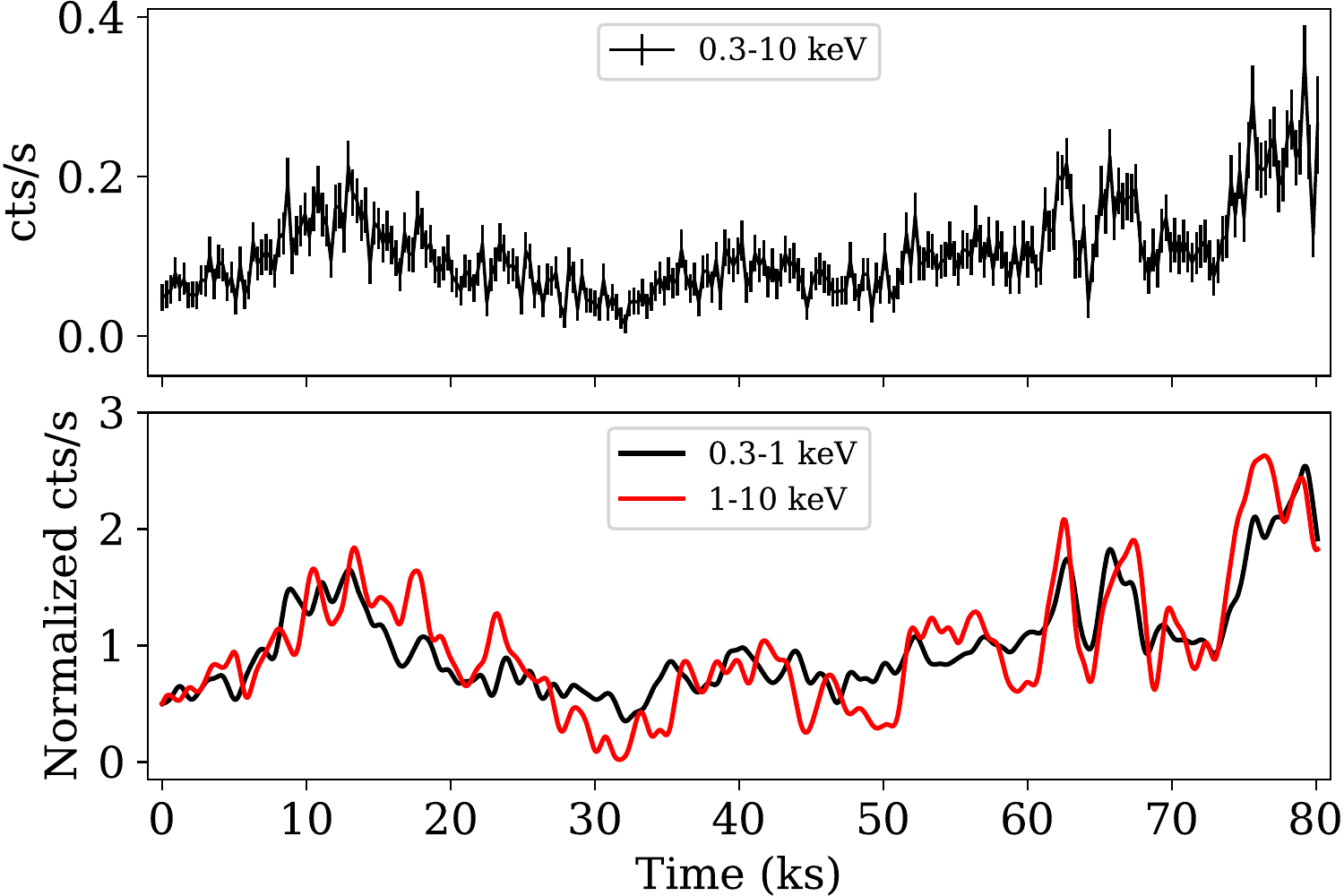}
    \caption{The top panel shows the background subtracted EPIC-pn
      lightcurve in the 0.3--10 keV energy band, with bin size
      ${\Delta}t=300$~s. A time shift between the peaks of the soft 
    (0.3--1 keV) and hard (1--10 keV) bands is visible in the bottom panel.}
    \label{fig:lc}
\end{figure}

NGC 7456 ULX-1  was recognized by \cite{pintore2020} as highly variable,
and here we perform a detailed exploration of it's variability. The
top panel of Fig.~\ref{fig:lc} shows the total band (0.3--10 keV)
lightcurve, where the  variability on time scales of ks is visible. 
The lightcurves of the selected soft (0.3--1 keV) and hard 
(1--10 keV) band are shown in the bottom panel of Fig.~\ref{fig:lc}.
These lightcurves have been smoothed using a Gaussian kernel with a
width of 500~s to reduce the random fluctuations due to Poisson
noise. One can see there is a slight delay between the peaks of the
two bands. 

\begin{figure}
    \centering
    \includegraphics[trim={0 0 0 0},width=0.35\textwidth]{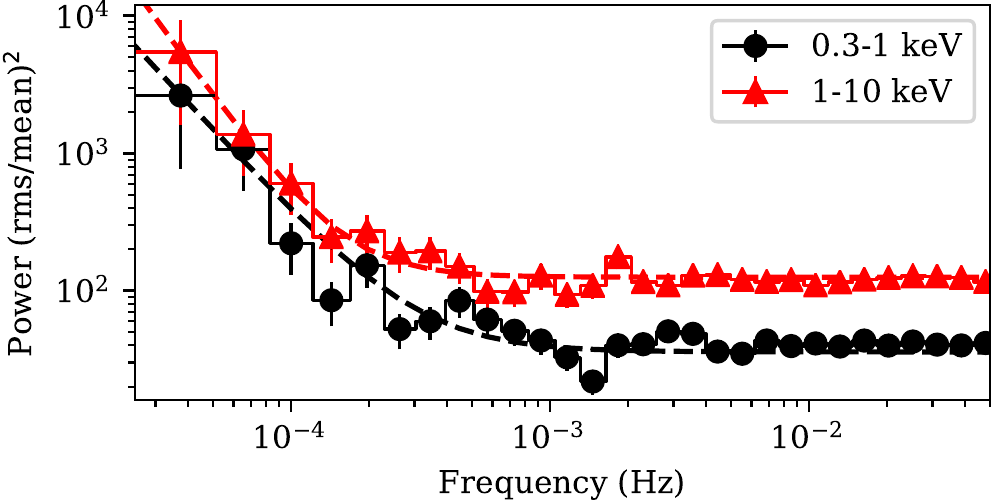}
    \includegraphics[trim={0 0 0 0},width=0.35\textwidth]{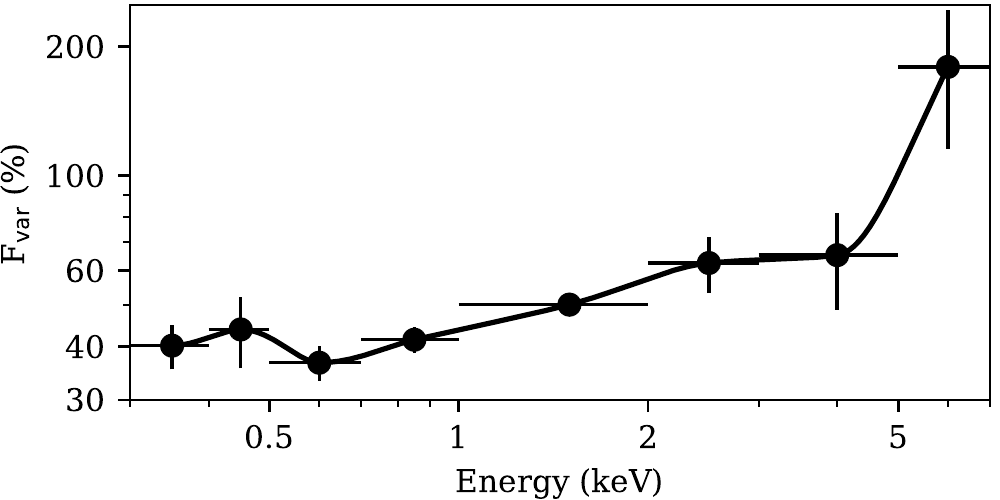}
    \caption{Top panel: PSDs for the soft and hard bands.
     The dashed lines show the best fits for a model composed of a simple power law plus a constant. Bottom panel: 
     fractional variability as a function of energy computed in the frequency interval $[0.25$--$5]\times10^{-4}$ Hz.
     The points have been connected with a monotonic cubic 
    interpolation to guide the eye.}
    \label{fig:PSD_fvar}
\end{figure}

As the lightcurves show enhanced variability, we performed Fourier timing
analysis to estimate power spectral density functions (PSD). The PSDs were
computed by averaging over duration of $\sim40$ ks and then logarathmically
re-binning by a factor of 1.24. The top panel of
Fig.~\ref{fig:PSD_fvar} shows the PSDs in fractional rms squared
normalization for both soft and hard
bands. There is no obvious visual evidence for a power-law break.
Therefore, we fitted the PSDs using a simple power law plus a
constant for the Poisson noise which dominated at high frequency:
$A(f/10^{-4}\rm Hz)^{-\beta}+C$, where $A$ is the power at $10^{-4}$
Hz.  The fit parameters, with 1$\sigma$ errors, for the soft band are
$A=359.56\pm30.11$, $\beta=2.03\pm0.09$, $C=35.82\pm9.84$; for the
hard band, $A=430.44\pm13.43$, $\beta=2.56\pm0.03$,
$C=152.33\pm4.81$. The Poisson noise starts to dominate above
$\sim$0.5 mHz. The PSD fitting reveals that the hard band has more
high-frequency variability power than the soft band at 0.1--0.3 mHz, which means the
hard band is more variable on short time scales than the soft band. This is
expected considering that hard photons come from compact regions closer to
the central object than soft photons, thus producing more rapid
variability.

To check if the variability really increases with energy we compute
the fractional variability, $F_{\rm var}$
\citep{edelson2002,vaughan2003}, from lightcurves in eight different
energy bands (0.3--0.4, 0.4--0.5, 0.5--0.7, 0.7--1,
1--2, 2--3, 3--5, and 5--7 keV). The signal to noise ratio
above 7 keV is too low for meaningful energy-dependent analysis constraints.
We used time bins of 1000~s because
the Poisson noise dominates on shorter timescales, corresponding to
frequencies $>0.5$ mHz (Fig. \ref{fig:PSD_fvar} upper panel).  This
allowed us to sample the red-noise dominated part of the PSD.
The lightcurves are chopped into two segments of 40 ks in length.
Then $F_{\rm var}$ was computed separately for each segment and
averaged over the two segments.
$F_{\rm var}$ thus gives the power spectra integrated over the frequency
between $2.5\times10^{-5}$ Hz to 0.5 mHz. The bottom panel of
Fig.~\ref{fig:PSD_fvar} shows $F_{\rm var}$ increasing as a
function of energy above 1 keV. $F_{\rm var}$ is almost constant
(within the errors) from 0.3 keV up to 1 keV. This may indicate that
softer photons come from a single emitting region.

\subsection{Time lag analysis}
\begin{figure}
    \centering
    \includegraphics[trim={0 0 0 0},width=0.35\textwidth]{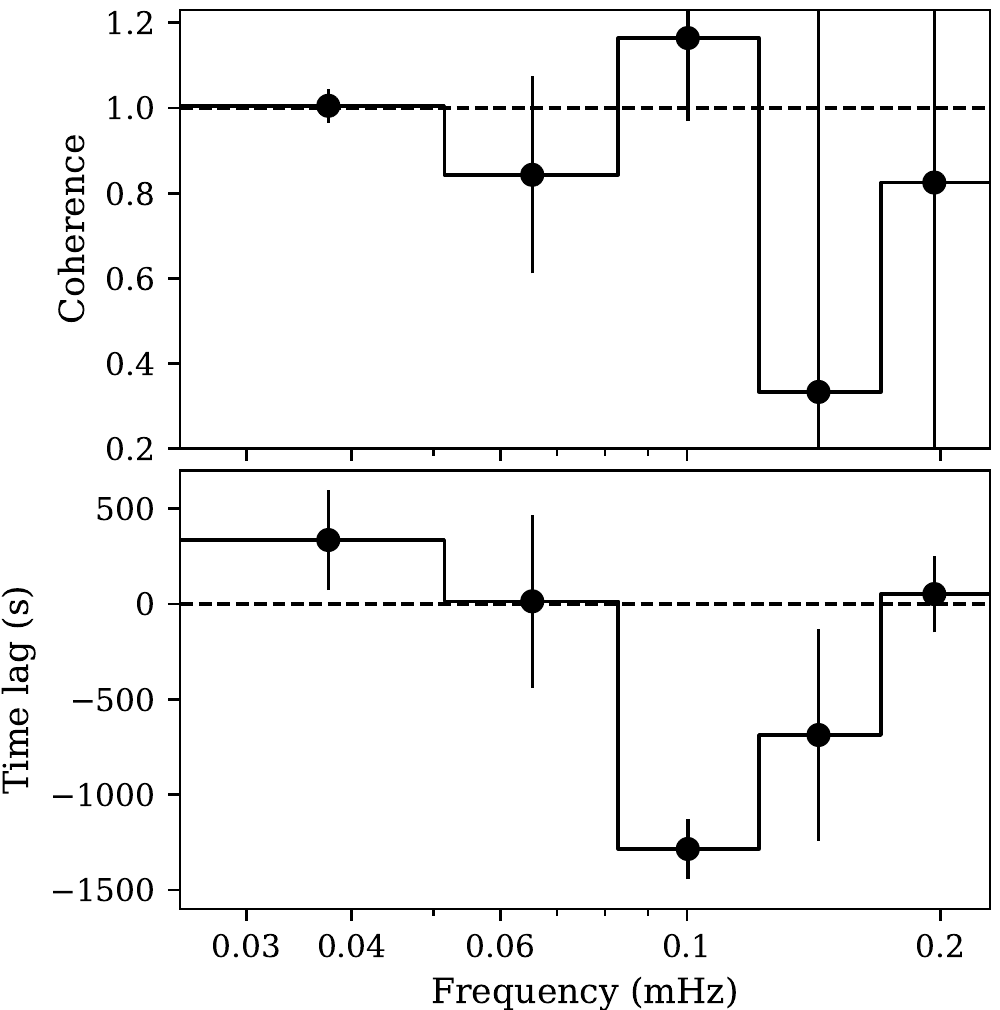}
    \caption{The coherence (top panel) and time lag (bottom panel) versus frequency 
    plot for the un-smoothed lightcurves. A negative lag means that soft photons lag hard photons. }
    \label{fig:lag_fr}
\end{figure}

To confirm and quantify the delay shown in Fig.~\ref{fig:lc}, we
measured the frequency-dependent time lag between the 
(un-smoothed) soft and the hard band lightcurves, following the
procedure outlined in \cite{uttley2014}. First we chopped the soft and
hard band lightcurves into two segments $\sim40$~ks long. Then we
computed the cross spectrum from each individual segment. Next, we
averaged the cross spectra over the two segments, and the averaged
cross spectrum was re-binned logarathmically using a factor of 1.24. The
time lag is given by the formula $\tau=\phi(f)/2\pi f$, where
$\phi(f)$ is the phase lag obtained from the re-binned averaged cross
spectrum. We used the modulus of the cross spectrum to compute the
coherence, a measurement of the level of linear correlation between
the two light curves \citep{vaughan1997}.  Fig.~\ref{fig:lag_fr} shows
the frequency-dependent coherence (top panel) and time lag (bottom
panel). The noise corrected coherence and its error are computed using
Eq.~8 in \cite{vaughan1997}.  The coherence is consistent with unity
after correction for the Poisson noise, meaning the variations in one
band are linearly well-correlated with variations in the other band
\citep{vaughan1997}. We follow the convention that the negative lag
means that the soft X-ray photons lag behind the hard X-rays and vice
versa for positive lag. We observe a negative lag at frequencies above
$\sim$0.07 mHz. At low frequencies lags are consistent with a hard lag
as we previously saw after smoothing the lightcurves with a Gaussian
kernel, and the amplitude of this lag is $336\pm263$~s in the frequency 
bin 0.025--0.05 mHz. To prove that the soft lag is not caused by 
the Poisson noise, we simulated lightcurves using the \cite{timmer1995} 
algorithm, generating coherent soft and hand band light curves with identical 
PSD properties, Poisson noise, and count rates to those measured for the 
real data. Only 14/1000 trials yielded lags more negative than $-$1300 s 
at 0.1 mHz, rejecting the notion of the lag being due to photon noise at 
$2.4\sigma$ confidence.

\begin{figure}
    \centering
    \includegraphics[trim={0 0 0 0},width=0.35\textwidth]{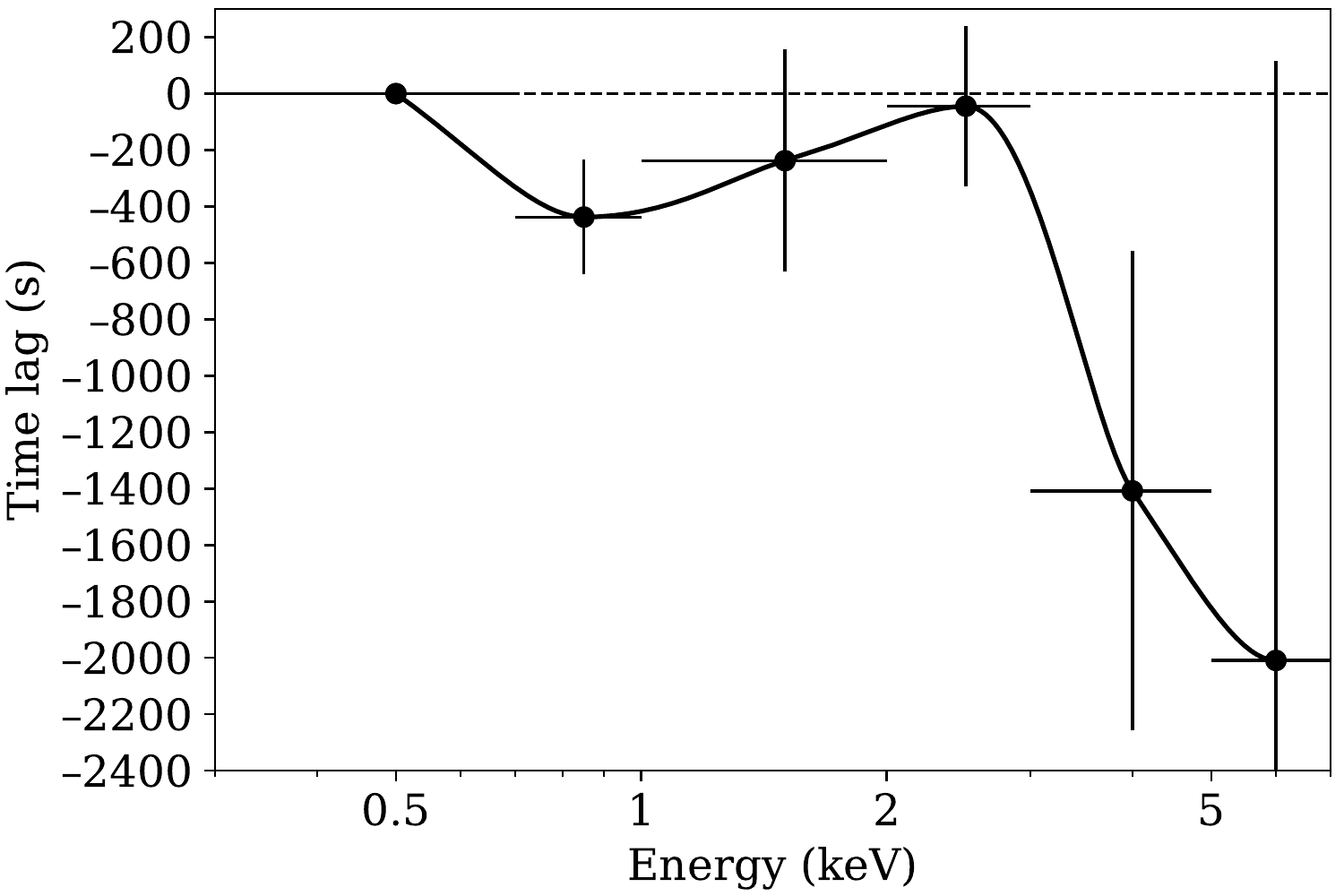}
    \caption{The lag-energy spectra in the frequency interval [0.7--1.2]$\times10^{-4}$ Hz. 
    The points have been connected with a monotonic cubic interpolation to guide the eye.}
    \label{fig:lag_en}
\end{figure}

Next, the lag energy spectrum was obtained (see Fig.~\ref{fig:lag_en})
by computing the cross spectra between the reference band and adjacent
energy bands. We used 0.3--0.7 keV as the reference band. Furthermore, the
lag was estimated from the resultant cross spectra which was averaged
over frequencies where we detected the soft lag with high coherence in
Fig.~\ref{fig:lag_fr}, roughly [0.7--1.2]$\times10^{-4}$ Hz.  The lags
have not been shifted, so zero lag means there is no time delay
between that band and the reference band. Similarly a negative lag
means the bin leads the reference band. The lag shows a sharp drop above
3 keV.

\begin{figure}
    \centering
    \includegraphics[trim={0 0 0 0},width=0.35\textwidth]{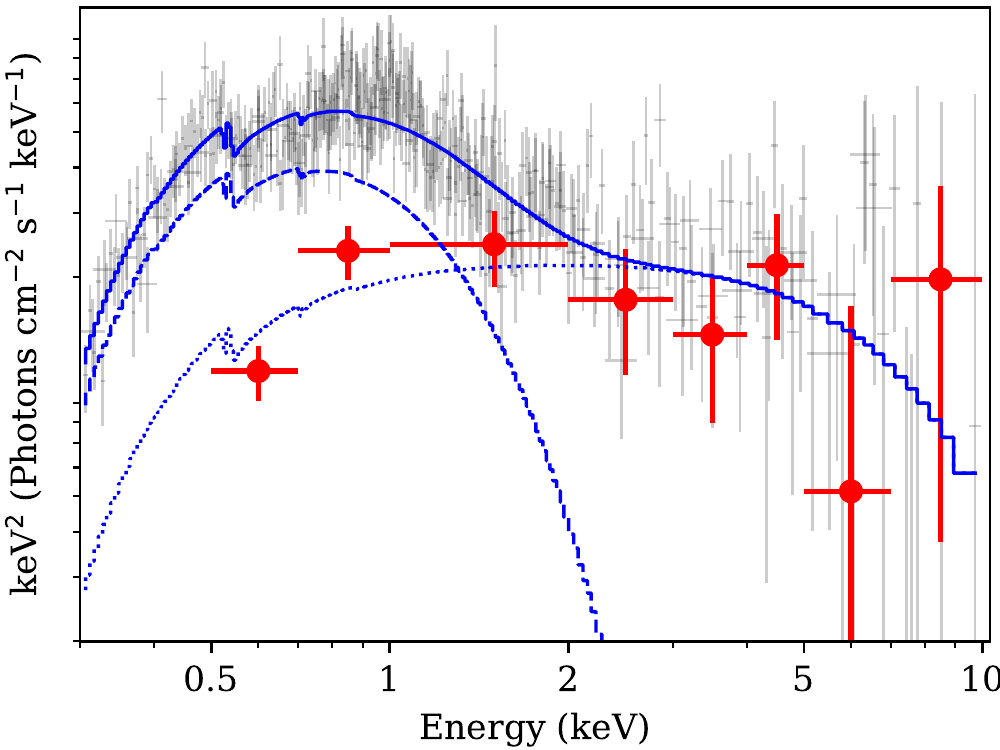}
    \caption{The figure shows the covariance spectrum (red data points) computed over the
      frequency range [0.5--1.5]$\times10^{-4}$ Hz, compared with the time-averaged energy spectrum 
     (light gray points). The Y-axis is scaled arbitrarily to compare the shape of the covariance spectra with the
      time-averaged spectrum. The covariance spectrum follows the shape 
    of the Comptonization component (dotted curve).}
    \label{fig:covari}
\end{figure}

We further computed the frequency-dependent covariance spectrum to
check which component of the energy spectrum is associated with the
correlated variability.  The covariance is computed following the
prescription outlined in \cite{wilkinson2009} in a similar frequency
interval (0.05--0.15 mHz)
as the lag-energy spectrum. Then the covariance spectrum was
loaded into {\sc xspec} and modeled with an absorbed power law
with column density fixed to $5.84\times10^{20}\ \rm cm^{-2}$,
obtained from the time averaged continuum fitting. The unfolded
covariance spectrum is shown in Fig.~\ref{fig:covari} (red points;
covariance is shifted along the Y axis) together with the time
averaged spectrum for comparison. It is apparent that the covariance
spectrum is harder than the time averaged spectrum and seems to follow
the shape of the \texttt{nthcomp} component. The covariance spectrum
provides independent confirmation about which spectral component is
responsible for the observed correlated variability.

\section{Discussion and conclusion}
We report the detection of an Fe K$_{\alpha}$ line in ULX-1 in
NGC~7456. The 0.3--10 keV mean spectrum is broadly described by a
two-component model where the soft component dominates in the 0.3--1
keV band and the hard component in the 1--10 keV band. The source is
in the ultraluminous soft state \citep{gladstone2009,sutton2013} in
which the soft component peaks over the hard component
(Fig.~\ref{fig:eeuf}).

The Fe$_{\alpha}$ line is detected at $2.6\sigma$ confidence, but the line width is poorly constrained, and we obtain an upper limit on
the width $\sigma<$300 eV. The full width at half-maximum of velocity
broadening is $v_{\rm FWHM}<$32900 km s$^{-1}$. In ULXs the geometry
of accreting gas is still unknown. %the possibility of super-Eddington accretion with geometrical beaming and an ultra-fast wind outflow makes the situation even more complicated. 
For simplicity, we will assume <$v^2$> =
$v^2_{\rm FWHM}$ in ULXs.  Then we use <$v^2$> = $\frac{GM}{R}$ to infer the
distance of the line-emitting gas
from the central compact object under the assumption that the
line-emitting gas is in Keplerian motion around the central compact
object. We obtain a distance of $R$ =
$\frac{c^2}{<v^2>}r_{\rm g}>85r_{\rm g}$.

So far, Fe K$_{\alpha}$ emission has not been detected in many ULXs,
and NGC 7456 ULX-1 would be  the second ULX source where
this line is detected.  Here, the line has an $EW$ of
$2000^{+1500}_{-1100}$ eV,
described in Sec.~\ref{sec:spec_fit}.
Previously, an Fe K$_{\alpha}$ line was detected in M82 X-1:
\cite{strohmayer2003} used an \emph{XMM-Newton} observation, and
reported an EW of 230--1300 eV depending on spectral fitting model;
an EW of 30--80 eV was reported by \cite{caballero-garcia2011}, who
used a \emph{Suzaku} observation.  Up to now, the iron line in NGC
7456 ULX-1 has the highest EW ever detected in ULX sources.
The data above 10 keV  with high spacial resolution 
is  necessary to confirm if the line originates in disk reflection; 
however, both the low flux  and ultraluminous soft spectral state of 
NGC 7456 ULX-1 make the source a poor candidate for a \emph{NuSTAR} observation.

NGC 7456 ULX-1 has 0.3--10 keV $F_{\rm var}=44.25\pm1.46\%$ over time scales of 1000 s to 40 ks.
This is the highest short term variability amplitude measured in ULXs so far. 
%So far ULXs with 0.3--10 keV $F_{\rm var}$ upto 10--20\% have been detected \citep{sutton2013}. 
%The increasing trend of variability amplitude as a function of energy in bottom panel of Fig. \ref{fig:PSD_fvar} 
%indicates strong spectral variability of the source. 
We found that the disk component varies little, but the emission above 1 keV is increasingly variable. 

Our timing analysis of NGC 7456 ULX-1 indicates the source is variable
on ks time scales, and we detected the soft X-ray band lagging behind the
hard X-ray band with a $\sim$1300 s delay at 0.1 mHz. The covariance
spectrum of NGC 7456 ULX-1 follows the shape of the harder spectral component and
there is a clear lack of a soft excess, similar to other
variable ULXs \citep{middleton2015}. The lack of variability in the
softer component suggests it plays no role in generating the soft
X-ray lag. Furthermore, the high coherence between the soft and hard bands
means there is a single driver of the variability.  Together
these are consistent with the soft lag being intrinsic to the harder
component and not being a delay between the two spectral
components. In the other words the harder component shows the
rapid variability and the low energy photons of this component arrive
over a ks later than the high energy photons.

\begin{figure}
   \centering
    \includegraphics[trim={0 0 0 0},width=0.40\textwidth]{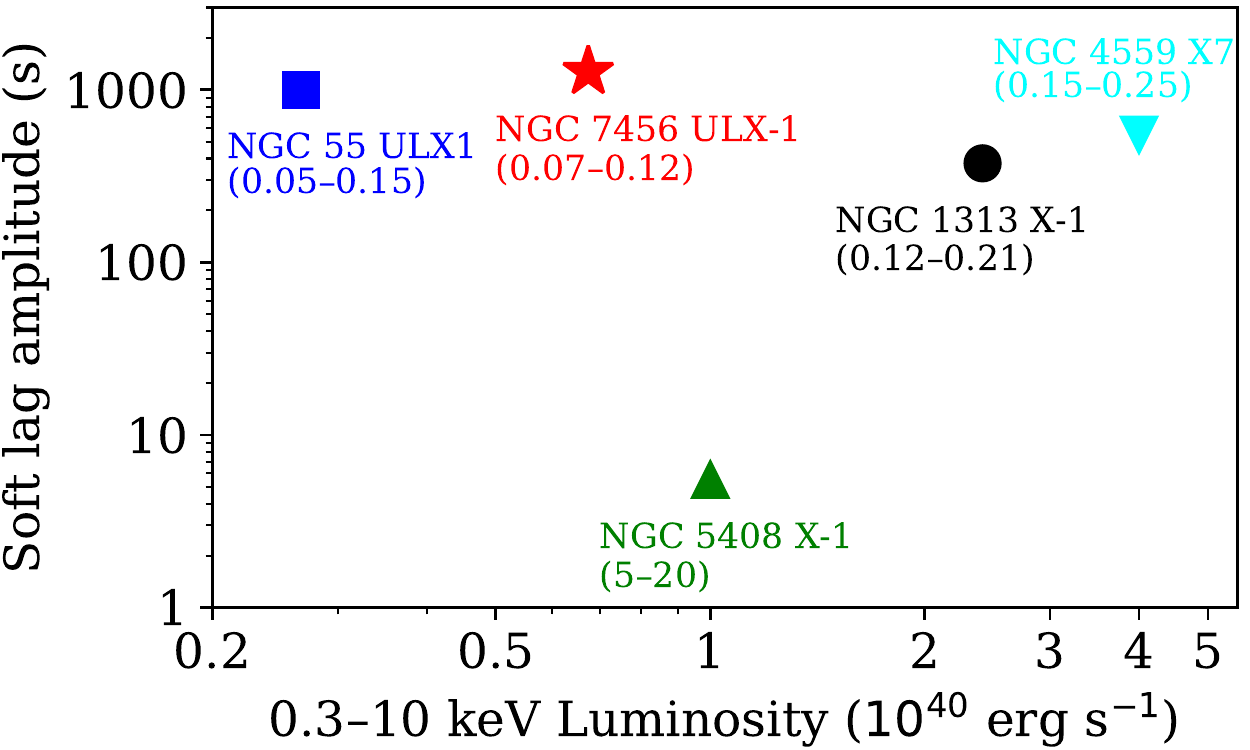}
   \caption{The lag amplitude of NGC 7456 ULX-1 compared with the lag amplitudes discovered in other ULXs:
      NGC 55 ULX1 \citep{pinto2017}, NGC 5408 X-1 \citep{demarco2013} NGC 1313 X-1 \citep{kara2020} and NGC 4559 X7 \citep{pintore2021}. 
      The number in parenthesis indicates the frequency range in mHz where the soft lag is detected.
    }
    \label{fig:lag_lumi}
\end{figure}

NGC 7456 ULX-1 would be the fifth ULX source to exhibit a soft lag. A
high frequency soft X-ray lag in a ULX was first detected in NGC 5408
X-1 by \cite{heil2010} and later by \cite{demarco2013} and
\cite{hernandez-garcia2015}. Later, a soft X-ray lag was detected in
NGC~55 ULX1 \citep{pinto2017}, NGC 1313 X-1 \citep{kara2020} and recently in NGC 4559 X7 \citep{pintore2021}. 
In Fig.~\ref{fig:lag_lumi} we plot the unabsorbed
0.3--10 keV X-ray luminosity versus soft lag amplitude of ULXs
detected in similar energy bands.
%0.2--0.3 keV vs 2--6 keV, 0.3--1 keV vs 1--10 keV, 0.3--2 keV vs 2--10 keV and 0.3--1 keV vs 1--7 keV for NGC 55 ULX1, NGC 7456 ULX-1, NGC 5408 X-1 and NGC 1313 X-1, respectively. 
NGC 55 ULX1 and NGC 7456 ULX-1 
show nearly the same lag amplitude in a
similar frequency interval. NGC 1313 X-1 and NGC 4559 X7 show a somewhat shorter lag,
but in a higher frequency interval. Given the similar amplitudes, it is possible that the
soft lag in these four sources have the same physical origin. On the
other hand, the lag amplitude in 5408 X-1 is an order of magnitude
smaller. The different behaviour of NGC 5408 X-1 may be due to the
soft lag in this source having a different origin from the other
ULXs. \cite{demarco2013} measured the soft X-ray lag in NGC 5408 X-1
with amplitude of $\sim$5 s on the time scale 5--20 mHz. Furthermore,
\cite{hernandez-garcia2015} extended the analysis to longer time
scales. The authors found a longer soft lag amplitude of $\sim$100 s
at $\sim$0.35 mHz but not long enough to be comparable to the other four 
ULXs. Moreover at even lower frequencies, around 0.1 mHz,
\cite{hernandez-garcia2015} observed a hard lag. So, it is clear that
NGC 5408 X-1 is an outlier, relative to the other four sources.

The soft X-ray lag in AGN is well established and the
explanation for its origin is thought to be due to light travel time
delays between the primary coronal emission and the reprocessed
emission (reverberation) in the accretion disk within roughly 
$10r_{\rm g}$ of the central BH \citep{demarco2013a}. If the lag
in NGC 7456 ULX-1 is really due to reverberation from the inner region
of the accretion disk, the $\sim$1300 s delay would imply a BH mass of
$\sim10^7\ \rm M_{\odot}$, which is extremely high. The other
explanation could be that reflection of the primary X-ray occurs at
larger distances from the central compact object. If we consider that
NGC 7456 ULX-1 is a stellar or intermediate mass BH of
$10$--$10^{4}\ \rm M_{\odot}$, then the reverberation must be
originating from $\sim10^{7}$--$10^{4}\ r_{\rm g}$, which is
consistent with a distant reflector.  Furthermore, the lack of soft
excess in the covariance spectra (see \citealt{uttley2011} for the
soft excess in covariance spectra of GX 339--4) makes the
reverberation model to be highly unlikely for possible explanation of
the soft X-ray lag origin in NGC 7456 ULX-1.

From variability studies of ULXs it has been found that soft ULXs show
more variability than hard ULXs and the variability is strongest at
higher energies \citep{sutton2013}.  This is often explained under the
scenario that the soft ULXs are seen through outflowing wind material
which intermittently blocks our line of sight to the hot central
region. Therefore the softer photons come from the down scattering of
hard photons. This leads to a higher flux in the softer band so the source
appears as a soft ULX and the intermittent blocking of hard photons
by the wind leads to large variability at higher energies. This picture
perfectly fits into the case of NGC 7456 ULX-1. The origin of the soft
lag could be explained in a similar picture. In such a picture the
absorption opacity of the material should be low for high energy
photons and high for low energy photons. So the low energy photons of
the harder component go through a large number of scatterings into
optically thick wind, compared to the high energy photons. This will
introduce a long delay between the arrival times of low and high energy
photons of the harder component and the lag amplitude will give the extent
of the outflow. However absorption may decrease the coherence
\cite[e.g.][]{demarco2020}, particularly if the wind has a high column
density and/or low ionization. On the contrary, we measure a high
coherence for NGC 7456 ULX-1 (Fig. \ref{fig:lag_fr}). %This model is not that simple and straightforward; making more quantitative estimates requires detailed modeling, which is beyond the scope of this paper.

\section{Acknowledgements}
AR was partially supported by Polish National Science Center grants
No. 2015/17/B/ST9/03422 and 2015/18/M/ST9/00541. BDM acknowledges support
from Ram{\'o}ny Cajal Fellowship RYC2018-025950-I.  AGM was partially
supported by Polish National Science Center grants 2016/23/B/ST9/03123
and 2018/31/G/ST9/03224. %\sm{This research has made use of a collection of ISIS functions (ISIS scripts) provided by ECAP/Remeis observatory and MIT.}

\section{Data Availability}
The data is publicly available from ESA's XMM-Newton Science Archive and 
NASA's HEASARC archive.

\bibliographystyle{mnras}
\bibliography{refs} % if your bibtex file is called example.bib

\begin{thebibliography}{}
\makeatletter
\relax
\def\mn@urlcharsother{\let\do\@makeother \do\$\do\&\do\#\do\^\do\_\do\%\do\~}
\def\mn@doi{\begingroup\mn@urlcharsother \@ifnextchar [ {\mn@doi@}
  {\mn@doi@[]}}
\def\mn@doi@[#1]#2{\def\@tempa{#1}\ifx\@tempa\@empty \href
  {http://dx.doi.org/#2} {doi:#2}\else \href {http://dx.doi.org/#2} {#1}\fi
  \endgroup}
\def\mn@eprint#1#2{\mn@eprint@#1:#2::\@nil}
\def\mn@eprint@arXiv#1{\href {http://arxiv.org/abs/#1} {{\tt arXiv:#1}}}
\def\mn@eprint@dblp#1{\href {http://dblp.uni-trier.de/rec/bibtex/#1.xml}
  {dblp:#1}}
\def\mn@eprint@#1:#2:#3:#4\@nil{\def\@tempa {#1}\def\@tempb {#2}\def\@tempc
  {#3}\ifx \@tempc \@empty \let \@tempc \@tempb \let \@tempb \@tempa \fi \ifx
  \@tempb \@empty \def\@tempb {arXiv}\fi \@ifundefined
  {mn@eprint@\@tempb}{\@tempb:\@tempc}{\expandafter \expandafter \csname
  mn@eprint@\@tempb\endcsname \expandafter{\@tempc}}}

\bibitem[\protect\citeauthoryear{{Akaike}}{{Akaike}}{1974}]{akaike1974}
{Akaike} H.,  1974, IEEE Transactions on Automatic Control, \href
  {https://ui.adsabs.harvard.edu/abs/1974ITAC...19..716A} {19, 716}

\bibitem[\protect\citeauthoryear{{Arnaud}}{{Arnaud}}{1996}]{arnaud1996}
{Arnaud} K.~A.,  1996, {XSPEC: The First Ten Years}.
p.~17

\bibitem[\protect\citeauthoryear{{Bachetti} et~al.,}{{Bachetti}
  et~al.}{2014}]{bachetti14}
{Bachetti} M.,  et~al., 2014, \mn@doi [\nat] {10.1038/nature13791}, \href
  {http://adsabs.harvard.edu/abs/2014Natur.514..202B} {514, 202}

\bibitem[\protect\citeauthoryear{{Brightman} et~al.,}{{Brightman}
  et~al.}{2018}]{brightman2018}
{Brightman} M.,  et~al., 2018, \mn@doi [Nature Astronomy]
  {10.1038/s41550-018-0391-6}, \href
  {https://ui.adsabs.harvard.edu/abs/2018NatAs...2..312B} {2, 312}

\bibitem[\protect\citeauthoryear{{Caballero-Garc{\'\i}a}}{{Caballero-Garc{\'\i}a}}{2011}]{caballero-garcia2011}
{Caballero-Garc{\'\i}a} M.~D.,  2011, \mn@doi [\mnras]
  {10.1111/j.1365-2966.2011.19615.x}, \href
  {https://ui.adsabs.harvard.edu/abs/2011MNRAS.418.1973C} {418, 1973}

\bibitem[\protect\citeauthoryear{{Carpano}, {Haberl}, {Maitra}  \&
  {Vasilopoulos}}{{Carpano} et~al.}{2018}]{carpano2018}
{Carpano} S.,  {Haberl} F.,  {Maitra} C.,   {Vasilopoulos} G.,  2018, \mn@doi
  [\mnras] {10.1093/mnrasl/sly030}, \href
  {https://ui.adsabs.harvard.edu/abs/2018MNRAS.476L..45C} {476, L45}

\bibitem[\protect\citeauthoryear{{De Marco}, {Ponti}, {Cappi}, {Dadina},
  {Uttley}, {Cackett}, {Fabian}  \& {Miniutti}}{{De Marco}
  et~al.}{2013a}]{demarco2013a}
{De Marco} B.,  {Ponti} G.,  {Cappi} M.,  {Dadina} M.,  {Uttley} P.,  {Cackett}
  E.~M.,  {Fabian} A.~C.,   {Miniutti} G.,  2013a, \mn@doi [\mnras]
  {10.1093/mnras/stt339}, \href
  {https://ui.adsabs.harvard.edu/abs/2013MNRAS.431.2441D} {431, 2441}

\bibitem[\protect\citeauthoryear{{De Marco}, {Ponti}, {Miniutti}, {Belloni},
  {Cappi}, {Dadina}  \& {Mu{\~n}oz-Darias}}{{De Marco}
  et~al.}{2013b}]{demarco2013}
{De Marco} B.,  {Ponti} G.,  {Miniutti} G.,  {Belloni} T.,  {Cappi} M.,
  {Dadina} M.,   {Mu{\~n}oz-Darias} T.,  2013b, \mn@doi [\mnras]
  {10.1093/mnras/stt1853}, \href
  {https://ui.adsabs.harvard.edu/abs/2013MNRAS.436.3782D} {436, 3782}

\bibitem[\protect\citeauthoryear{{De Marco} et~al.,}{{De Marco}
  et~al.}{2020}]{demarco2020}
{De Marco} B.,  et~al., 2020, \mn@doi [\aap] {10.1051/0004-6361/201936470},
  \href {https://ui.adsabs.harvard.edu/abs/2020A&A...634A..65D} {634, A65}

\bibitem[\protect\citeauthoryear{{Earnshaw} et~al.,}{{Earnshaw}
  et~al.}{2016}]{earnshaw2016}
{Earnshaw} H.~M.,  et~al., 2016, \mn@doi [\mnras] {10.1093/mnras/stv2945},
  \href {https://ui.adsabs.harvard.edu/abs/2016MNRAS.456.3840E} {456, 3840}

\bibitem[\protect\citeauthoryear{{Edelson}, {Turner}, {Pounds}, {Vaughan},
  {Markowitz}, {Marshall}, {Dobbie}  \& {Warwick}}{{Edelson}
  et~al.}{2002}]{edelson2002}
{Edelson} R.,  {Turner} T.~J.,  {Pounds} K.,  {Vaughan} S.,  {Markowitz} A.,
  {Marshall} H.,  {Dobbie} P.,   {Warwick} R.,  2002, \mn@doi [\apj]
  {10.1086/323779}, \href
  {https://ui.adsabs.harvard.edu/abs/2002ApJ...568..610E} {568, 610}

\bibitem[\protect\citeauthoryear{{F{\"u}rst} et~al.,}{{F{\"u}rst}
  et~al.}{2016}]{furst2016}
{F{\"u}rst} F.,  et~al., 2016, \mn@doi [\apjl] {10.3847/2041-8205/831/2/L14},
  \href {http://adsabs.harvard.edu/abs/2016ApJ...831L..14F} {831, L14}

\bibitem[\protect\citeauthoryear{{Gladstone}, {Roberts}  \& {Done}}{{Gladstone}
  et~al.}{2009}]{gladstone2009}
{Gladstone} J.~C.,  {Roberts} T.~P.,   {Done} C.,  2009, \mn@doi [\mnras]
  {10.1111/j.1365-2966.2009.15123.x}, \href
  {https://ui.adsabs.harvard.edu/abs/2009MNRAS.397.1836G} {397, 1836}

\bibitem[\protect\citeauthoryear{{Heil} \& {Vaughan}}{{Heil} \&
  {Vaughan}}{2010}]{heil2010}
{Heil} L.~M.,  {Vaughan} S.,  2010, \mn@doi [\mnras]
  {10.1111/j.1745-3933.2010.00864.x}, \href
  {https://ui.adsabs.harvard.edu/abs/2010MNRAS.405L..86H} {405, L86}

\bibitem[\protect\citeauthoryear{{Heil}, {Vaughan}  \& {Roberts}}{{Heil}
  et~al.}{2009}]{heil2009}
{Heil} L.~M.,  {Vaughan} S.,   {Roberts} T.~P.,  2009, \mn@doi [\mnras]
  {10.1111/j.1365-2966.2009.15068.x}, \href
  {https://ui.adsabs.harvard.edu/abs/2009MNRAS.397.1061H} {397, 1061}

\bibitem[\protect\citeauthoryear{{Hern{\'a}ndez-Garc{\'\i}a}, {Vaughan},
  {Roberts}  \& {Middleton}}{{Hern{\'a}ndez-Garc{\'\i}a}
  et~al.}{2015}]{hernandez-garcia2015}
{Hern{\'a}ndez-Garc{\'\i}a} L.,  {Vaughan} S.,  {Roberts} T.~P.,   {Middleton}
  M.,  2015, \mn@doi [\mnras] {10.1093/mnras/stv1830}, \href
  {https://ui.adsabs.harvard.edu/abs/2015MNRAS.453.2877H} {453, 2877}

\bibitem[\protect\citeauthoryear{{Houck} \& {Denicola}}{{Houck} \&
  {Denicola}}{2000}]{houck2000}
{Houck} J.~C.,  {Denicola} L.~A.,  2000, in {Manset} N.,  {Veillet} C.,
  {Crabtree} D.,  eds,  Astronomical Society of the Pacific Conference Series
  Vol. 216, Astronomical Data Analysis Software and Systems IX. p.~591

\bibitem[\protect\citeauthoryear{{Israel} et~al.,}{{Israel}
  et~al.}{2017a}]{israel2017a}
{Israel} G.~L.,  et~al., 2017a, \mn@doi [Science] {10.1126/science.aai8635},
  \href {http://adsabs.harvard.edu/abs/2017Sci...355..817I} {355, 817}

\bibitem[\protect\citeauthoryear{{Israel} et~al.,}{{Israel}
  et~al.}{2017b}]{israel2017b}
{Israel} G.~L.,  et~al., 2017b, \mn@doi [\mnras] {10.1093/mnrasl/slw218}, \href
  {http://adsabs.harvard.edu/abs/2017MNRAS.466L..48I} {466, L48}

\bibitem[\protect\citeauthoryear{{Jansen} et~al.,}{{Jansen}
  et~al.}{2001}]{jansen2001}
{Jansen} F.,  et~al., 2001, \mn@doi [\aap] {10.1051/0004-6361:20000036}, \href
  {http://adsabs.harvard.edu/abs/2001A%26A...365L...1J} {365, L1}

\bibitem[\protect\citeauthoryear{{Kaastra} \& {Bleeker}}{{Kaastra} \&
  {Bleeker}}{2016}]{kaastra2016}
{Kaastra} J.~S.,  {Bleeker} J.~A.~M.,  2016, \mn@doi [\aap]
  {10.1051/0004-6361/201527395}, \href
  {https://ui.adsabs.harvard.edu/abs/2016A&A...587A.151K} {587, A151}

\bibitem[\protect\citeauthoryear{{Kara} et~al.,}{{Kara}
  et~al.}{2020}]{kara2020}
{Kara} E.,  et~al., 2020, \mn@doi [\mnras] {10.1093/mnras/stz3318}, \href
  {https://ui.adsabs.harvard.edu/abs/2020MNRAS.491.5172K} {491, 5172}

\bibitem[\protect\citeauthoryear{{Middleton}, {Heil}, {Pintore}, {Walton}  \&
  {Roberts}}{{Middleton} et~al.}{2015}]{middleton2015}
{Middleton} M.~J.,  {Heil} L.,  {Pintore} F.,  {Walton} D.~J.,   {Roberts}
  T.~P.,  2015, \mn@doi [\mnras] {10.1093/mnras/stu2644}, \href
  {https://ui.adsabs.harvard.edu/abs/2015MNRAS.447.3243M} {447, 3243}

\bibitem[\protect\citeauthoryear{{Mondal}, {R{\'o}{\.z}a{\'n}ska}, {Lai}  \&
  {De Marco}}{{Mondal} et~al.}{2020}]{mondal2020b}
{Mondal} S.,  {R{\'o}{\.z}a{\'n}ska} A.,  {Lai} E.~V.,   {De Marco} B.,  2020,
  \mn@doi [\aap] {10.1051/0004-6361/202038684}, \href
  {https://ui.adsabs.harvard.edu/abs/2020A&A...642A..94M} {642, A94}

\bibitem[\protect\citeauthoryear{{Mondal}, {Rozanska}, {Baginska}, {Markowitz}
  \& {De Marco}}{{Mondal} et~al.}{2021}]{mondal2021}
{Mondal} S.,  {Rozanska} A.,  {Baginska} P.,  {Markowitz} A.,   {De Marco} B.,
  2021, arXiv e-prints, \href
  {https://ui.adsabs.harvard.edu/abs/2021arXiv210412894M} {p. arXiv:2104.12894}

\bibitem[\protect\citeauthoryear{{Pinto} et~al.,}{{Pinto}
  et~al.}{2017}]{pinto2017}
{Pinto} C.,  et~al., 2017, \mn@doi [\mnras] {10.1093/mnras/stx641}, \href
  {https://ui.adsabs.harvard.edu/abs/2017MNRAS.468.2865P} {468, 2865}

\bibitem[\protect\citeauthoryear{{Pintore} et~al.,}{{Pintore}
  et~al.}{2020}]{pintore2020}
{Pintore} F.,  et~al., 2020, \mn@doi [\apj] {10.3847/1538-4357/ab6ffd}, \href
  {https://ui.adsabs.harvard.edu/abs/2020ApJ...890..166P} {890, 166}

\bibitem[\protect\citeauthoryear{{Pintore} et~al.,}{{Pintore}
  et~al.}{2021}]{pintore2021}
{Pintore} F.,  et~al., 2021, \mn@doi [\mnras] {10.1093/mnras/stab913}, \href
  {https://ui.adsabs.harvard.edu/abs/2021MNRAS.tmp..933P} {}

\bibitem[\protect\citeauthoryear{{Rodr{\'\i}guez Castillo}
  et~al.,}{{Rodr{\'\i}guez Castillo} et~al.}{2020}]{rodriguez-castillo2020}
{Rodr{\'\i}guez Castillo} G.~A.,  et~al., 2020, \mn@doi [\apj]
  {10.3847/1538-4357/ab8a44}, \href
  {https://ui.adsabs.harvard.edu/abs/2020ApJ...895...60R} {895, 60}

\bibitem[\protect\citeauthoryear{{Sathyaprakash} et~al.,}{{Sathyaprakash}
  et~al.}{2019}]{sathyaprakash2019}
{Sathyaprakash} R.,  et~al., 2019, \mn@doi [\mnras] {10.1093/mnrasl/slz086},
  \href {https://ui.adsabs.harvard.edu/abs/2019MNRAS.488L..35S} {488, L35}

\bibitem[\protect\citeauthoryear{{Schwarz}}{{Schwarz}}{1978}]{schwarz1978}
{Schwarz} G.,  1978, Annals of Statistics, \href
  {https://ui.adsabs.harvard.edu/abs/1978AnSta...6..461S} {6, 461}

\bibitem[\protect\citeauthoryear{{Strohmayer} \& {Mushotzky}}{{Strohmayer} \&
  {Mushotzky}}{2003}]{strohmayer2003}
{Strohmayer} T.~E.,  {Mushotzky} R.~F.,  2003, \mn@doi [\apjl]
  {10.1086/374732}, \href
  {https://ui.adsabs.harvard.edu/abs/2003ApJ...586L..61S} {586, L61}

\bibitem[\protect\citeauthoryear{{Sutton}, {Roberts}  \& {Middleton}}{{Sutton}
  et~al.}{2013}]{sutton2013}
{Sutton} A.~D.,  {Roberts} T.~P.,   {Middleton} M.~J.,  2013, \mn@doi [\mnras]
  {10.1093/mnras/stt1419}, \href
  {https://ui.adsabs.harvard.edu/abs/2013MNRAS.435.1758S} {435, 1758}

\bibitem[\protect\citeauthoryear{{Timmer} \& {Koenig}}{{Timmer} \&
  {Koenig}}{1995}]{timmer1995}
{Timmer} J.,  {Koenig} M.,  1995, \aap, \href
  {https://ui.adsabs.harvard.edu/abs/1995A&A...300..707T} {300, 707}

\bibitem[\protect\citeauthoryear{{Tully}, {Courtois}  \& {Sorce}}{{Tully}
  et~al.}{2016}]{tully2016a}
{Tully} R.~B.,  {Courtois} H.~M.,   {Sorce} J.~G.,  2016, \mn@doi [\aj]
  {10.3847/0004-6256/152/2/50}, \href
  {https://ui.adsabs.harvard.edu/abs/2016AJ....152...50T} {152, 50}

\bibitem[\protect\citeauthoryear{{Uttley}, {Wilkinson}, {Cassatella}, {Wilms},
  {Pottschmidt}, {Hanke}  \& {B{\"o}ck}}{{Uttley} et~al.}{2011}]{uttley2011}
{Uttley} P.,  {Wilkinson} T.,  {Cassatella} P.,  {Wilms} J.,  {Pottschmidt} K.,
   {Hanke} M.,   {B{\"o}ck} M.,  2011, \mn@doi [\mnras]
  {10.1111/j.1745-3933.2011.01056.x}, \href
  {https://ui.adsabs.harvard.edu/abs/2011MNRAS.414L..60U} {414, L60}

\bibitem[\protect\citeauthoryear{{Uttley}, {Cackett}, {Fabian}, {Kara}  \&
  {Wilkins}}{{Uttley} et~al.}{2014}]{uttley2014}
{Uttley} P.,  {Cackett} E.~M.,  {Fabian} A.~C.,  {Kara} E.,   {Wilkins} D.~R.,
  2014, \mn@doi [\aapr] {10.1007/s00159-014-0072-0}, \href
  {https://ui.adsabs.harvard.edu/abs/2014A&ARv..22...72U} {22, 72}

\bibitem[\protect\citeauthoryear{{Vaughan} \& {Nowak}}{{Vaughan} \&
  {Nowak}}{1997}]{vaughan1997}
{Vaughan} B.~A.,  {Nowak} M.~A.,  1997, \mn@doi [\apjl] {10.1086/310430}, \href
  {https://ui.adsabs.harvard.edu/abs/1997ApJ...474L..43V} {474, L43}

\bibitem[\protect\citeauthoryear{{Vaughan}, {Edelson}, {Warwick}  \&
  {Uttley}}{{Vaughan} et~al.}{2003}]{vaughan2003}
{Vaughan} S.,  {Edelson} R.,  {Warwick} R.~S.,   {Uttley} P.,  2003, \mn@doi
  [\mnras] {10.1046/j.1365-2966.2003.07042.x}, \href
  {https://ui.adsabs.harvard.edu/abs/2003MNRAS.345.1271V} {345, 1271}

\bibitem[\protect\citeauthoryear{{Walton}, {Roberts}, {Mateos}  \&
  {Heard}}{{Walton} et~al.}{2011}]{walton2011}
{Walton} D.~J.,  {Roberts} T.~P.,  {Mateos} S.,   {Heard} V.,  2011, \mn@doi
  [\mnras] {10.1111/j.1365-2966.2011.19154.x}, \href
  {https://ui.adsabs.harvard.edu/abs/2011MNRAS.416.1844W} {416, 1844}

\bibitem[\protect\citeauthoryear{{Wilkinson} \& {Uttley}}{{Wilkinson} \&
  {Uttley}}{2009}]{wilkinson2009}
{Wilkinson} T.,  {Uttley} P.,  2009, \mn@doi [\mnras]
  {10.1111/j.1365-2966.2009.15008.x}, \href
  {https://ui.adsabs.harvard.edu/abs/2009MNRAS.397..666W} {397, 666}

\makeatother
\end{thebibliography}

% Alternatively you could enter them by hand, like this:
% This method is tedious and prone to error if you have lots of references
%\begin{thebibliography}{99}
%\bibitem[\protect\citeauthoryear{Author}{2012}]{Author2012}
%Author A.~N., 2013, Journal of Improbable Astronomy, 1, 1
%\bibitem[\protect\citeauthoryear{Others}{2013}]{Others2013}
%Others S., 2012, Journal of Interesting Stuff, 17, 198
%\end{thebibliography}

%%%%%%%%%%%%%%%%%%%%%%%%%%%%%%%%%%%%%%%%%%%%%%%%%%
\end{document}